\documentclass[conference]{IEEEtran}

\usepackage{graphicx}
\usepackage{subfig}
\usepackage{epstopdf}
\usepackage{bm}

\usepackage{floatrow}
\begin{document}
%%% CHANGES FROM 1. SUBMITTED PAPER (FOR REVIEW)
%%% update of related work -> highlight multi carrier in the section (3?) and scope of our work (In our work we want to prove the performance of that technique, but allow only a single antenna mode. Further, we do not use any expensive measurement equipment, but more prototype like receivers such as Software Defined Radio platforms to estimate the wireless channel and transmit signals. Instead of different antenna modes, we use the samples of the frequency domain channel estimation as our feature space.

\title{Physical Layer Authentication for Mission Critical Machine Type Communication using Gaussian Mixture Model based Clustering}
%Physical Layer Authentication for Mission Critical Machine Type Communication based on Gaussian Mixture Models and Carrier Frequency Offset estimation

%\author{{Andreas Weinand, Michael Karrenbauer and Hans D. Schotten} \\
	%Institute for Wireless Communication and Navigation\\
	%University of Kaiserslautern, Germany\\
	%Email: \{weinand, karrenbauer, schotten\}@eit.uni-kl.de
\author{\IEEEauthorblockN{Andreas Weinand, Michael Karrenbauer, Ji Lianghai, Hans D. Schotten}
\IEEEauthorblockA{Institute for Wireless Communication and Navigation\\
University of Kaiserslautern, Germany\\
Email: \{weinand, karrenbauer, ji, schotten\}@eit.uni-kl.de
}
%\and
%\IEEEauthorblockN{Abhijit Ambekar}
%\IEEEauthorblockA{German Research Center for Artificial Intelligence (DFKI)\\
%Kaiserslautern, Germany\\
%Email: abhijit.ambekar@dfki.de}
%\and
%\IEEEauthorblockN{James Kirk\\ and Montgomery Scott}
%\IEEEauthorblockA{Starfleet Academy\\
%San Francisco, California 96678--2391\\
%Fax: (888) 555--1212}
}


%Telephone: (800) 555--1212, Fax: (888) 555--1212}


\maketitle

\begin{abstract}
The application of Mission Critical Machine Type Communication (MC-MTC) in wireless systems is currently a hot research topic. Wireless systems are considered to provide numerous advantages over wired systems in e.g. industrial applications such as closed loop control. However, due to the broadcast nature of the wireless channel, such systems are prone to a wide range of cyber attacks. These range from passive eavesdropping attacks to active attacks like data manipulation or masquerade attacks. Therefore it is necessary to provide reliable and efficient security mechanisms. Some of the most important security issues in such a system are to ensure integrity as well as authenticity of exchanged messages over the air between communicating devices. In the present work, an approach on how to achieve this goal in MC-MTC systems based on Physical Layer Security (PHYSEC) is presented. A new method that clusters channel estimates of different transmitters based on a Gaussian Mixture Model is applied for that purpose. Further, an experimental proof-of-concept evaluation is given and we compare the performance of our approach with a mean square error based detection method. 
\end{abstract}

% \fi
\IEEEpeerreviewmaketitle

\section{Introduction}
\label{intro}

{\let\thefootnote\relax\footnote{This is a preprint, the full paper has been published in Proceedings of 2017 IEEE 85th Vehicular Technology Conference (VTC2017-Spring), \copyright 2017 IEEE. Personal use of this material is permitted. However, permission to use this material for any other purposes must be obtained from the IEEE by sending a request to pubs-permissions@ieee.org.}}
\IEEEpubidadjcol

Recently, a new trend in the area of wireless systems is the operation of MC-MTC as for instance closed loop control applications. These have much higher requirements regarding reliability, availability and especially latency compared to common applications such as media streaming or web browsing over IEEE 802.11 based wireless systems or today's cellular systems. Another important requirement in the area of MC-MTC is the fact that secure transmission of data has to be taken into account. Due to the sensitive information transmitted in e. g. industrial or automotive scenarios, it is necessary to guarantee a high degree of information security. Especially authenticity as well as integrity of the transmitted data has to be ensured to prohibit a wide range of possible active cyber attacks. For this purpose identification and authentication of received messages is necessary in order to be sure of the originator of that data before it is consumed and processed by the respective application.\\
% IETF Richtlinie MAC Länge, Verfahren, HMAC/CMAC
Although there are conventional cryptography techniques to ensure authenticity as well as integrity of message payload, these require a lot of resources. Especially they lead to increase in message size due to the fact that, for example message authentication codes (MAC) which are used in IEEE 802.15.4 based systems, add a kind of check sum to the actual message payload. In IEEE 802.15.4 based systems either a 4, 8 or 16 Byte MAC is added to the respective payload. The recommendation of the IETF is to either use a CMAC, e. g. based on AES-128 block cipher, or a HMAC which is based on a cryptographic hash function. For AES-128 based CMAC a maximum shortening to 64 Bit is recommended, while for HMAC a minimum MAC size of 80 Bit is recommended. If we now assume that the payload of a MC-MTC packet has a length of 32 Byte (the dimension of this assumption is e.g. confirmed by \cite{Osman.2015}), then the payload overhead regarding the MAC size for AES-128 CMAC of 8 Byte would be already $20\%$.  %%% IETF Länge 
This means that a latency overhead of at least $20\%$ is added by only applying means to guarantee message authenticity and integrity. Another important issue is, that key based schemes such as message authentication codes are only able to protect the message payload from the mentioned attacks. An attacker is still able to perform attacks such as address spoofing, or even worse, record a message and replay it after a while. Due to these drawbacks, another idea is to check for authenticity of a message at a lower level by taking physical properties of the radio link signal in time domain, as well as in spatial domain into account. In this work, a keyless approach for this based on estimating the wireless channel at link level is presented. As mentioned, MC-MTC and closed loop control applications are considered here, which in combination seems to be a perfect case for our approach, as we can assume that frequent and periodic data transmissions and with this channel estimation at the same rate is carried out. For experimental evaluation, we consider an OFDM system and based on the respective frequency domain channel estimations, we decide from which source a received data packet was transmitted.  
\\
%%% organisation of the paper
The remainder of the work is organized as follows. In section \ref{related work} we give a short overview on related work with respect to previous considered approaches and in section \ref{system_model} we describe the system model. Our approach of Gaussian Mixture Model based clustering is presented in section \ref{approach}. In section \ref{results} we present the results of our work and section \ref{CONC} finally concludes the paper. 

\section{Related Work}
\label{related work}
Several approaches on exploiting the wireless channel for security purposes, also known as PHYSEC, have been investigated recently. In \cite{Jorswieck.2015} a good overview on this topic is given. While many works have focused on extracting secret keys between two communicating devices, such as \cite{Guillaume.}, \cite{Zenger.2014}, \cite{Ambekar.2014}, the focus of our work is on guaranteeing secure transmission with respect to authenticity of data packets from one device to another. One of the first works considering that idea has been for example \cite{Xiao.2007}, where an approach based on simulation of the wireless channel and hypothesis testing is presented for static scenarios and is later in \cite{Xiao.2008} extended to time-variant scenarios. In \cite{Pei.2014}, two approaches based on machine learning, Support Vector Machine and Linear Fisher Discriminant Analysis, are presented. The approach considered in \cite{Tugnait.2010} is similar to our approach, as they propose a CSI-based authentication method for a single carrier system. The second approach considered in \cite{Tugnait.2010} is whiteness of residuals testing. In \cite{Shi.2013} an RSS-based approach for body area networks is presented. The work in \cite{Refaey.2014} considers a multilayer approach based on OFDM to guarantee authentication of TCP packets. A Gaussian Mixture Model based technique in combination with exploitation of the channel responses for different antenna modes is considered in \cite{Gulati.2013}. 
\section{System model}
\label{system_model}

In this section we describe the system model and the attacker model including the mentioned active attacks. Further, the channel model is introduced and we explain how to exploit PHYSEC techniques, actually frequent channel estimation, in order to overcome these attack scenarios. 

\subsection{Attacker model}
\begin{figure}[t]
\centering
\includegraphics[width=0.5\textwidth]{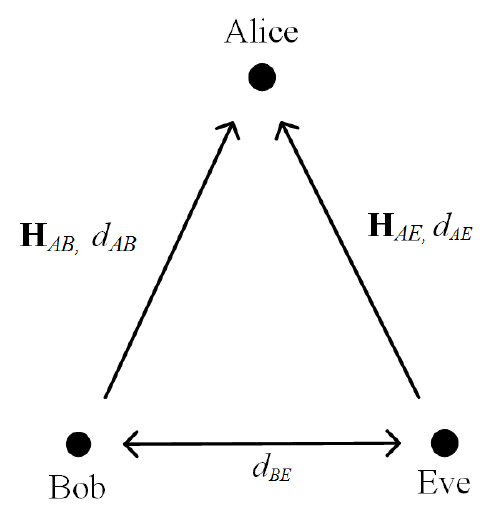}
\caption{System and Attacker Model}
\label{attacker_model}
\end{figure}

We consider two users, Alice and Bob, who want to exchange authenticated messages with each other. For this work we define that Bob is the legal transmit node who wishes to send some sensitive information to the legal receiver node Alice. Alice must make sure that Bob is the true transmitter of these messages. A third party Eve tries to masquerade as Bob and sends messages to Alice as well (see Fig. \ref{attacker_model}). A typical scenario for an attacker Eve is that he is at a spatially different location compared to Bob and uses advanced equipment such as directed antennas and high sensitivity receivers to maximize his range to his benefit. We also assume perfect knowledge of the underlying communication protocol at Eve to run active attacks such as masquerade attacks, replay attacks or address spoofing attacks. It is not assumed that Eve is gaining physical access to Alice or Bob to accomplish invasive attacks such as hardware modification. Further, other active attacks such as Denial-of-Service attacks due to jamming are not considered as well. It is assumed, that the legal communicating participants Bob and Alice have already carried out initial user authentication to each other and have set up trust in a secure way. Attacks on the initial authentication stage are not considered. The goal is now to authenticate the messages transmitted from Bob to Alice in a secure way, which as well takes the requirements of MC-MTC into account, especially minimization of transmission latency.\\

\subsection{Channel model and channel estimation}
%%% drawbacks in intro
%%% analytic channel model from literature
Due to the mentioned drawbacks of message authenticity checking based on conventional cryptography, such as message authentication codes, and the special requirements of MC-MTC, a more promising approach is to use characteristics of the wireless channel and the physical layer to decide about the origin of a received message. In our work, we focus on channel estimations which are computed at the receiver in any OFDM based system to perform for example channel equalization. In contrast to PHYSEC techniques such as secret key generation, which are based on the assumption that there is a lot of temporal variation in the wireless channel, our approach relies on the fact that the wireless channel does not vary significantly during subsequent channel measurements. However, the same idea that yields for both is to make use of the advantage of the fast spatial decorrelation property of wireless channels. For our work in particular, this means, that e.g. Alice receives messages from the legal transmit node Bob and estimates the actual channel $\mathbf{\hat{H}}$ as 
\begin{equation}
\mathbf{H}=[|h_1|,\ldots, |h_M|]
\label{eq:}
\end{equation}
with $|h_l|$ being the magnitude of the gain of the $l$-th subcarrier and $l=1,\ldots,M$.
%Actually, the channel estimated by Alice contains several different parts. Beside the real channel response in frequency domain $\mathbf{\hat{H}}=[{\hat{h}}_1,\ldots, {\hat{h}}_M]$ due to e.g. multipath propagation or doppler shifts, also the frequency characteristics of the transmit filter $\mathbf{H_{T}}=[{h}_{T,1},\ldots, {h}_{T,M}]$ as well as of the receive filter $\mathbf{H_{R}}=[{h}_{R,1},\ldots, {h}_{R,M}]$ are jointly estimated.
%We denote all these influences together as 
%\begin{equation}
%\mathbf{H}=\mathbf{H_{T}}\cdot \mathbf{\hat{H}}\cdot \mathbf{H_{R}} 
%\label{eq:}
%\end{equation}
%of a transmitter and receiver pair.
Due to the receiver noise figure or thermal noise, channel estimation is not perfect, which means that errors occur ($\mathbf{H}\ne\mathbf{\hat{H}}$). These influences can be modelled as a gaussian random variable $\mathbf{N}$ with zero mean and variance $\mathbf{\sigma_N^2}$. The transmit signal $\mathbf{X}$ will be received as
\begin{equation}
\mathbf{Y}=\mathbf{\hat{H}}\cdot \mathbf{X} + \mathbf{N}
\label{eq:}
\end{equation}
and consequently the channel is estimated as
\begin{equation}
\mathbf{H}=\mathbf{\hat{H}}+\frac{\mathbf{N}}{\mathbf{X}}.
\label{eq:}
\end{equation}
The idea is now that if an attacker Eve tries to transmit messages masqueraded as Bob, the channel measured by Alice is another one compared to the channel regarding to Bob. Basically, if we denote the channel estimates with $\mathbf{H}_{\mbox{\scriptsize AB}}$ and $\mathbf{H}_{\mbox{\scriptsize AE}}$ taken due to Bob and Eve respectively, this yields $\mathbf{H}_{\mbox{\scriptsize AE}} \ne \mathbf{H}_{\mbox{\scriptsize AB}}$. Due to the distance $d_{\mbox{\scriptsize BE}}$ between the attacker node Eve and the legal transmitter node Bob (which needs to be more than the wavelength of the transmitted signal), Eve is not able to masquerade without further effort due to the mentioned fast spatial decorrelation property of the channel. We can now use the estimated channel conditions by Alice to identify the originator of the respective message by means of clustering based on a Gaussian Mixture Model.

\section{Gaussian Mixture Model for PHYSEC based Authentication}
\label{approach}
This section deals with Gaussian Mixture Models which are used to cluster the channel estimates in combination with the EM algorithm. The result of this process is then used to make a decision about the corresponding transmitter of received data packets. 
\subsection{Gaussian Mixture Model}
A mixture of Gaussians 
\begin{equation}
f(\mathbf{x})=\sum_{k=1}^{K}\pi_k \mathcal{N}(\mathbf{x}|\bm{\mu}_k,\mathbf{\Sigma}_k)
\label{eq:}
\end{equation}
is consisting of $K$ Gaussian densities $f_k(\mathbf{x})=\mathcal{N}(\mathbf{x}|\bm{\mu}_k,\mathbf{\Sigma}_k)$ which each have a mean $\bm{\mu}_k$ and a covariance matrix $\mathbf{\Sigma}_k$. The mixtures are weighted by mixing coefficients $\bm{\pi}=\{\pi_1,\ldots,\pi_K\}$ which are normalized yielding
\begin{equation}
\sum_{k=1}^K\pi_k=1.
\label{eq:}
\end{equation}
These mixing coefficients further fulfill the property to be probabilities, technically they are prior probabilities. The goal is now to calculate the posterior probability
\begin{equation}
p_{i,k}^{(j)}=\frac{\pi_k^{(j)}\mathcal{N}(x_i|\mu_k^{(j)},\Sigma_k^{(j)})}{\sum_{k=1}^K\pi_k^{(j)}\mathcal{N}(x_i|\mu_k^{(j)},\Sigma_k^{(j)})}
\label{eq:}
\end{equation}
of each new set of data points $x_i$ with $i=1,\ldots,N$, i. e. each new channel estimate, which denotes the likelihood of this data belonging to a certain component of the mixture. The posterior probability is updated during the expectation step of the EM algorithm. In the maximization step the parameter values are updated. The weighting coefficients are calculated as
\begin{equation}
\pi_k^{(j+1)}=\frac{\sum_{i=1}^N p_{i,k}^{(j)}}{N},
\label{eq:}
\end{equation}
whereas the updated mean and covariance values are calculated as
\begin{equation}
\mu_k^{(j+1)}=\frac{\sum_{i=1}^N p_{i,k}^{(j)}x_i}{\sum_{i=1}^N p_{i,k}^{(j)}}
\label{eq:}
\end{equation}
and
\begin{equation}
\Sigma_k^{(j+1)}=\frac{\sum_{i=1}^N p_{i,k}^{(j)}(x_i-\mu_{k}^{(j)})(x_i-\mu_{k}^{(j)})^T}{\sum_{i=1}^N p_{i,k}^{(j)}}
\label{eq:}
\end{equation}
for the $j$-th iteration of the EM algorithm respectively. 

\subsection{Physical Layer Authentication based on Clustering}
To make a decision on received data packets from any transmitter, we need to determine how likely it is that a new set of data belongs to one of the gaussian mixture components. In our case we have $K=2$ mixture components, one for Bob and one for Eve each. If the channel estimate fits to the cluster modeling Bob, then Alice will assume that Bob is the true transmitter. She further can use this new information to update the gaussian mixture model, which improves the accuracy of it. Due to temporal variations, if e.g. one or more users have some degree of mobility, it is even necessary to continuously update the model online after a certain time to catch up with these variations. If the likelihood of belonging to Bobs cluster of a new set of data is below a certain threshold, then Alice assumes that it was introduced by Eve. 
In order to build an initial model and help Alice to identify who is belonging to which cluster, Bob needs to send some training messages to Alice. Alice will then use the cluster component with the most data sets from this training phase as the cluster belonging to Bob. To now attack our system, we assume that a message is either send to Alice by Bob or Eve by a given probability respectively. The probability of Eve transmitting (masqueraded as Bob) is also known as the attack intensity ($AI$). For each message, we decide about the originator based on the current model. After $N$ received messages, the model is updated based on these $N$ new data sets and the current GMM properties $(\bm{\pi},\bm{\mu},\mathbf{\Sigma})_m$ ($m$ denotes the index of the blocks of data sets) yielding the new model $(\bm{\pi},\bm{\mu},\mathbf{\Sigma})_{m+1}$. By doing this, the historical data that was used to build the model initially does not need to be stored. As a result of this, we get two performance parameters, the detection probability $P_{\mbox{\scriptsize D}}$ and the false alarm rate $P_{\mbox{\scriptsize FA}}$.
\begin{equation}
P_{\mbox{\scriptsize D}}=p_{i,B}((\pi_{B},\bm{\mu}_{B},\mathbf{\Sigma}_{B})_m|\mathbf{H}_{m,i} \mathrm{\ due\ to\ Eve}) < th
\label{eq:}
\end{equation}
denotes the probability of detecting Eve as the transmitter of the $i$-th message of the $m$-th set of messages under the condition that it was truly sent by Eve and
\begin{equation}
P_{\mbox{\scriptsize FA}}=p_{i,B}((\pi_{B},\bm{\mu}_{B},\mathbf{\Sigma}_{B})_{m}|\mathbf{H}_{m,i} \mathrm{\ due\ to\ Bob}) < th
\label{eq:}
\end{equation}
the probability of detecting Eve as transmitter of that message under the condition that it was truly sent by Bob.
%\begin{equation}
%%\[
     %\mathrm{Receive\ packet} \left\{\begin{array}{ll} \mathrm{process\ it}, & \mathrm{if}\ p(\mathbf{H}_i|\pi_{B},\bm{\mu}_{B},\mathbf{\Sigma}_{B}) \geq th \\
         %\mathrm{deny\ it}, & \mathrm{if}\ p(\mathbf{H}_i|\pi_{B},\bm{\mu}_{B},\mathbf{\Sigma}_{B}) < th\end{array}\right.
  %%\]
%\end{equation}
%In the first case of this equation, Alice assumes that there has been no manipulation attack performed on that respective packet and that it was send by the legal transmitter node Bob. If the probability of $\mathbf{H}_i$ belonging to the cluster of Bob is below a certain threshold $th$, Alice assumes that Eve was the originator of it. In our case, the threshold is applied to the posterior probability. \\

\section{Results}
\label{results}
%physical layer authentication, channel based authentication, channel profile monitoring-> nicht
%MC-MTC, URLLC
%todo: 
%abstract
%this paper: focus on direct update method
%intro: überarbeiten auf focus ausrichten
%related work: Referenzen überprüfen, chen k means wenn vergleich kmeans mit GMM
%channel profiles mathematische modellierung
%theoretisches (analytisches) Kanalmodell+transceiver Eigenschaften am Frequenzgang, realer Kanal vs. geschätzter Kanal Nomenklatur
%Kombination channel profiles und GMM zur Entscheidung + Update, Hypothesis formulation
%Kanalmessungen: BW, # of training messages, attack percentage, used frequency samples, frequency of channel measurements
%Auswertung: Modell mit t trainingsmessungen bilden, Entscheidung für neues Paket + Modell update, zufällig eingestreute Attacken?
%attack intensity: M packete jeweils nacheinander, Bob p%, Eve (p-1)%
%ROC curves: Posterior Probability threshold variieren, log wenn Werte sehr nah bei 0 oder 1
%\begin{figure*}[t!]
%\centering
%\subfloat[Case I]{\includegraphics[width=0.5\textwidth]{Results/MSE_time}%
%\hfil
%\subfloat[Case II]{\includegraphics[width=0.5\textwidth]{Results/MSE_space}%
%\label{fig_second_case}}
%\label{fig_sim}
%\vskip\baselineskip
%\subfloat[Case III]{\includegraphics[width=0.5\textwidth]{Results/PCC_time}%
%\label{fig_first_case2}}
%\hfil
%\subfloat[Case IV]{\includegraphics[width=0.5\textwidth]{Results/PCC_space}%
%\label{fig_second_case2}}
%\caption{Simulation results for the network.}
%\label{fig_sim}
%\end{figure*}
In this section we describe our setup for the experimental evaluation and show the final results of our work. 
\subsection{Experimental setup}
\begin{figure}[t]
\centering
\includegraphics[width=0.5\textwidth]{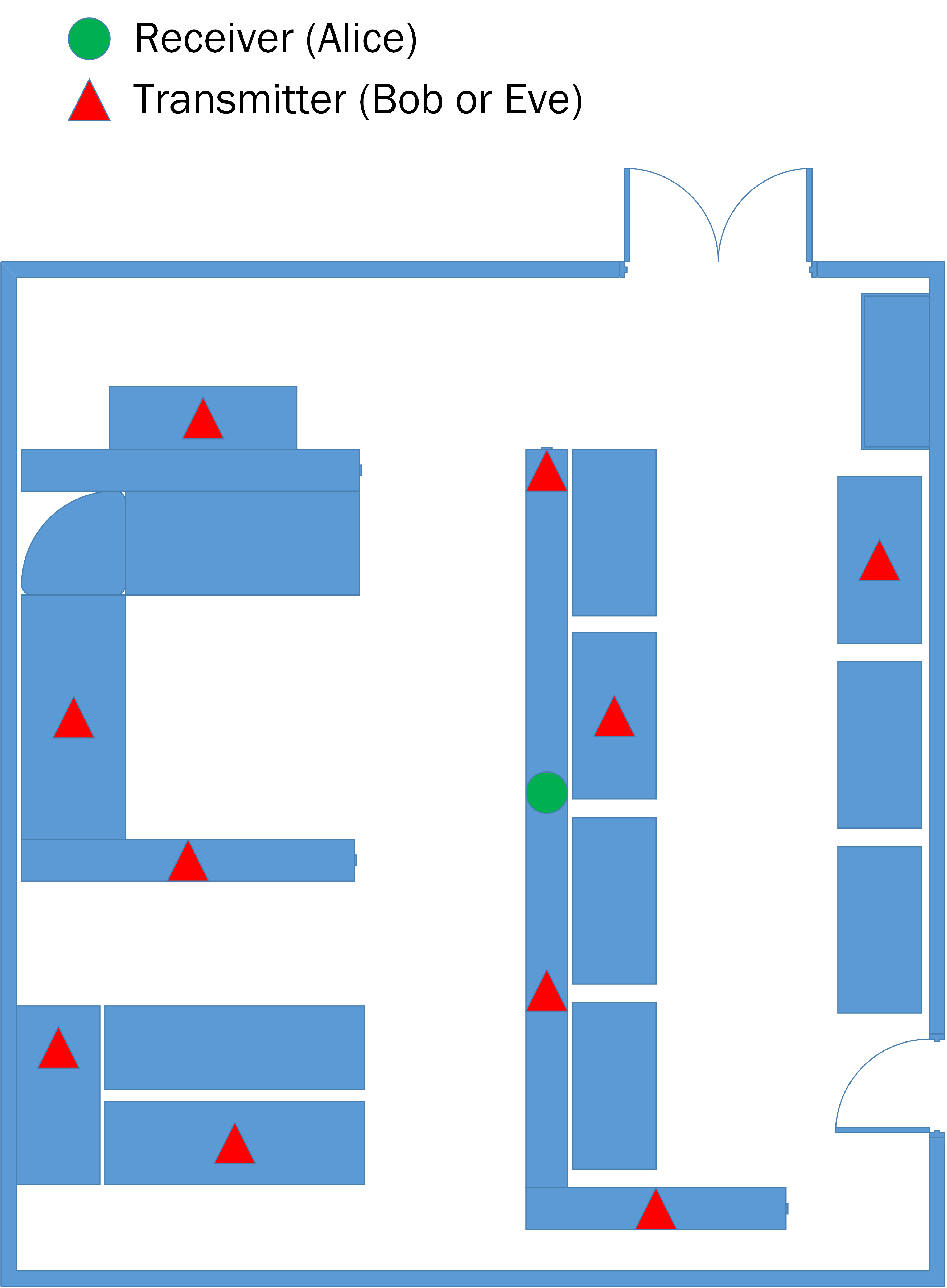}
\caption{Environment with different Alice and Bob/Eve positions}
\label{environment}
\end{figure}

%%% Werte anpassen je nachdem was wirklich genommen wurde
To evaluate our concepts, we use USRP N210 SDR platforms from Ettus Research with SBX daughterboards. We use GNURadio OFDM transmitter and receiver blocks to process data packets and perform channel estimation on each received data packet. A setup with an FFT size of $64$ is considered and $48$ active subcarriers. The cyclic prefix length is $16$ samples at a baseband sample rate of $3.125$ MSps, whereas the carrier frequency is $2.45$ GHz. For each received data packet, the initial channel taps are calculated based on the known Schmidl and Cox preamble \cite{Schmidl.1997} which is also used to calculate the frequency offset at the receiver (actually this preamble consists of two OFDM symbols). In each message, this preamble is followed by $37$ data symbols yielding a time resolution of 998.4 $\mu$s for the channel estimations. As a first step, we consider a static setup where all participants do not move during transmitting and receiving. The environment is a mixed office/lab area with a lot of objects and metal walls. Due to this we assume at least some amount of multipath propagation existing and with this frequency selective channels. We record data for several different locations of Bob and Eve respectively, yielding multiple different constellations of Alice/Bob and Alice/Eve pairs as shown in Fig. \ref{environment}. 
\subsection{Performance of GMM based Clustering}
\begin{figure}[t!]
\centering
\subfloat[ROC curves in linear scale]{\includegraphics[width=\textwidth]{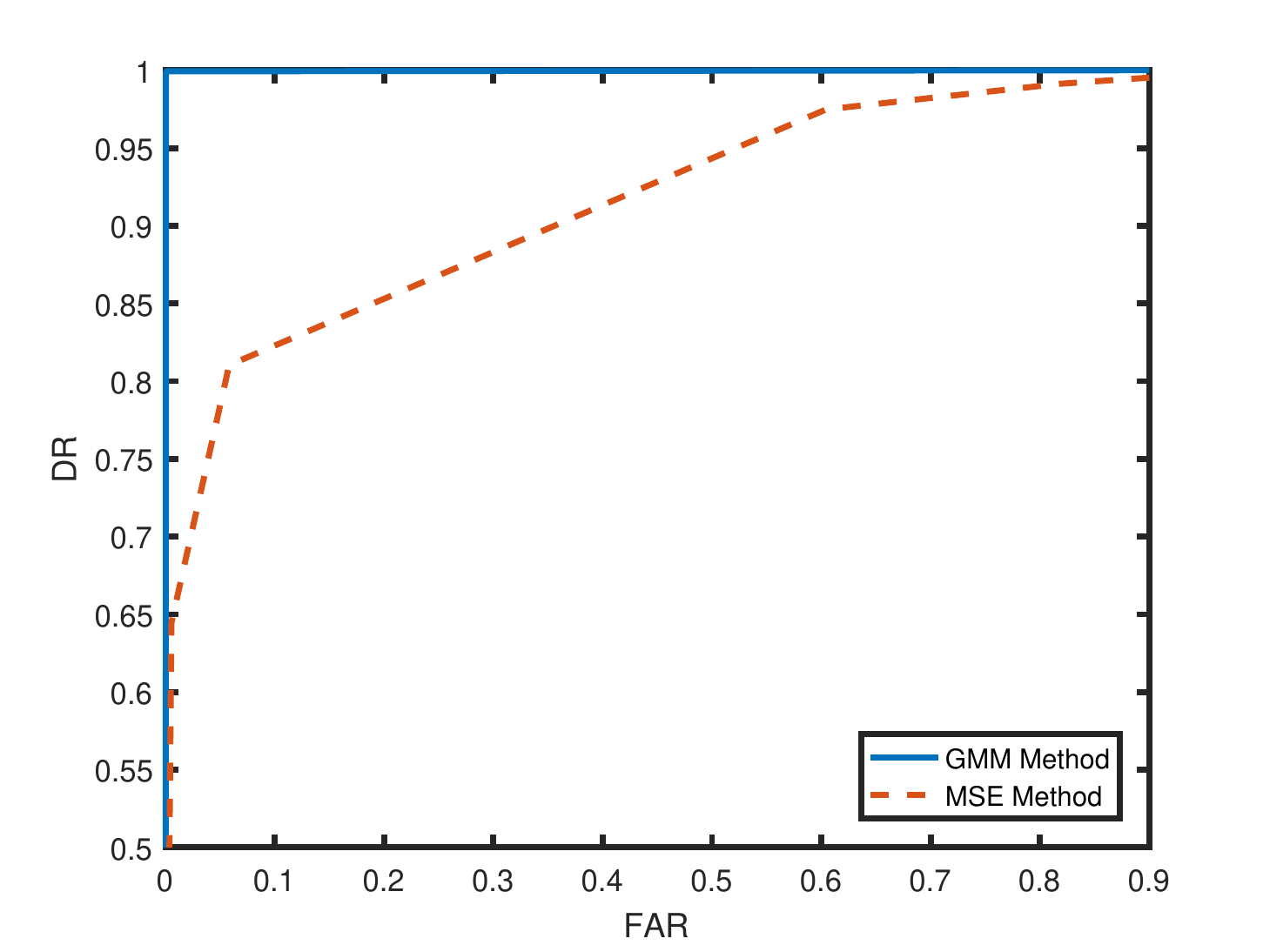}%
\label{roc_methods_1}}
\vfil
\subfloat[ROC curve in logarithmic scale]{\includegraphics[width=\textwidth]{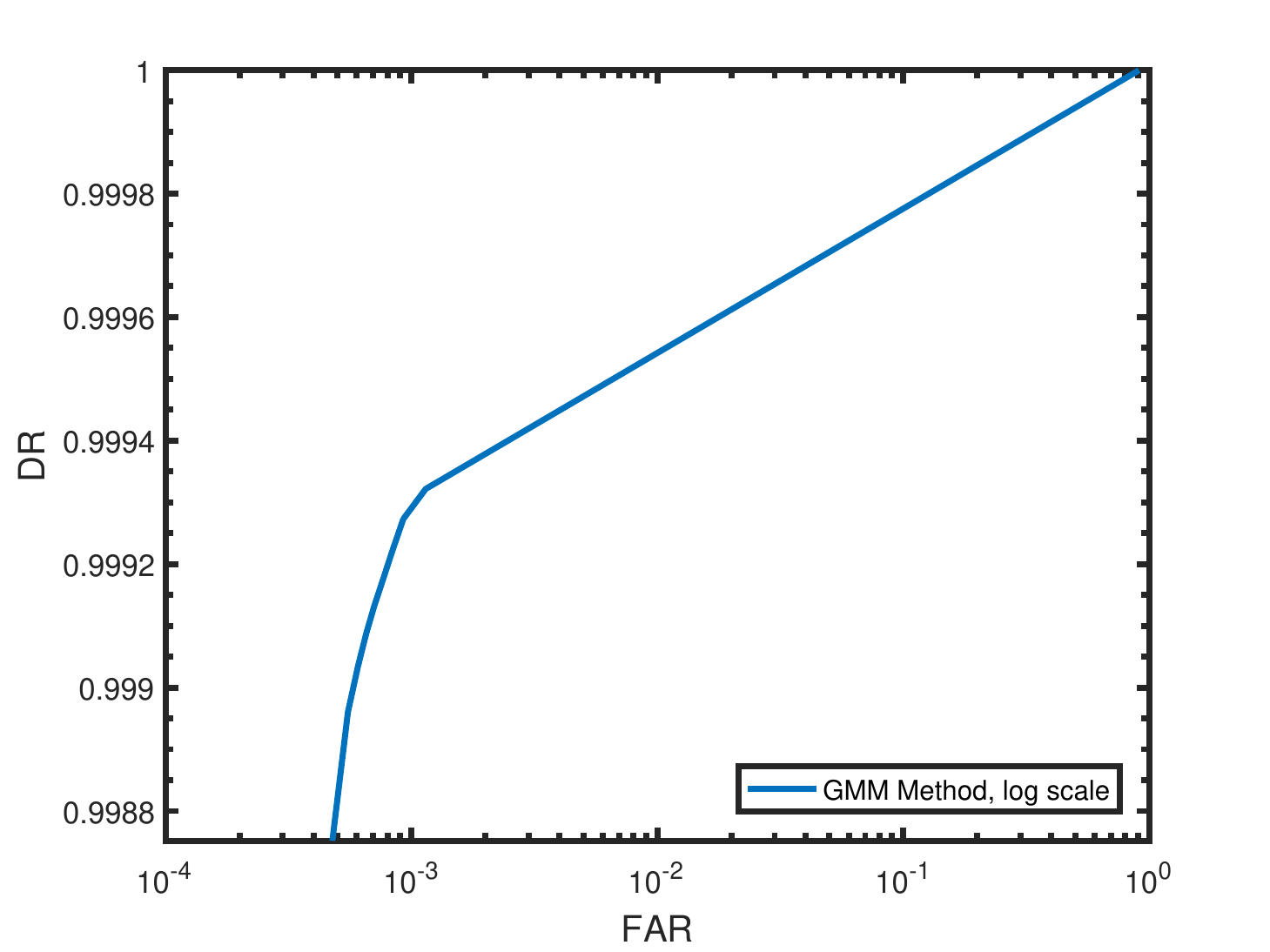}%
\label{roc_methods_2}}
\caption{ROC curves for the GMM and MSE method}
\label{roc_methods}
\end{figure}

% values need to be updated and figures
In order to evaluate the performance of the GMM based clustering, the detection probability $P_{\mbox{\scriptsize D}}$ as well as the false alarm rate $P_{\mbox{\scriptsize FA}}$ are considered and plotted in form of a ROC curve, where each data point is a pair of $P_{\mbox{\scriptsize D}}$ and $P_{\mbox{\scriptsize FA}}$ values at a certain threshold. We considered a block size of $N=1000$ data sets in order to update our GMM and use one block for training the model and $99$ data blocks in order to test it. The attack intensity is kept at $50\%$. Fig. \ref{roc_methods_1} shows the performance of our GMM based method compared to the simpler method of mean square error (MSE) based detection considered in \cite{Weinand.2016} for the case, that all $48$ active carriers are used for the detection. It can be seen that the GMM method outperforms the MSE based method. While the detection rate of the MSE method is $81.03\%$ at a false alarm rate of $5.83\%$, the detection rate of the GMM based method is $99.97\%$ (if cases of ill-conditioned covariance matrices are avoided) at the same false alarm rate. Even at a false alarm rate of $0.1\%$, the GMM method has a detection rate of $99.93\%$.

\begin{figure}[h!]
\centering
\subfloat[ROC curves in linear scale]{\includegraphics[width=\textwidth]{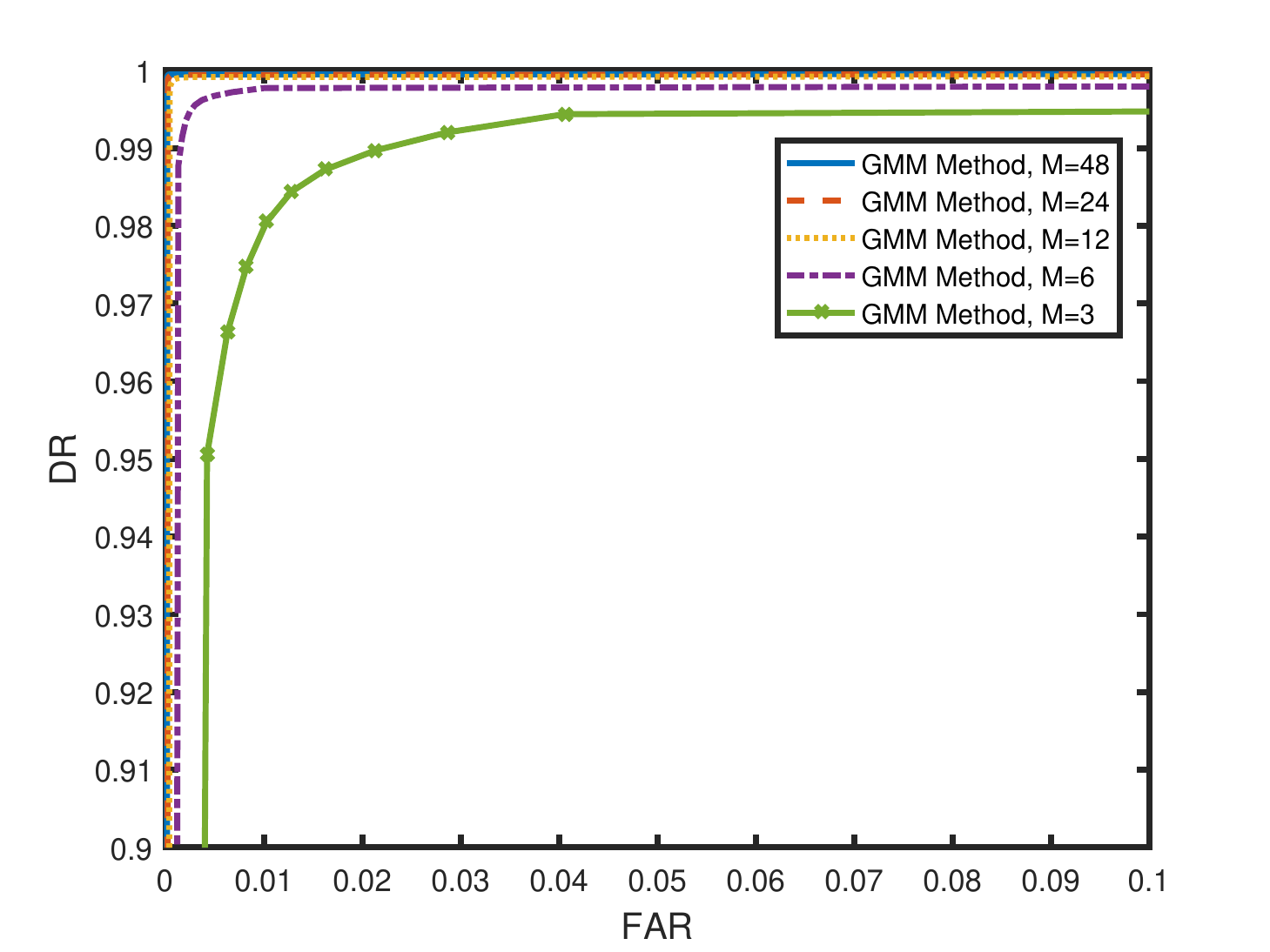}%
\label{roc_M_1}}
\vfil
\subfloat[ROC curves in logarithmic scale]{\includegraphics[width=\textwidth]{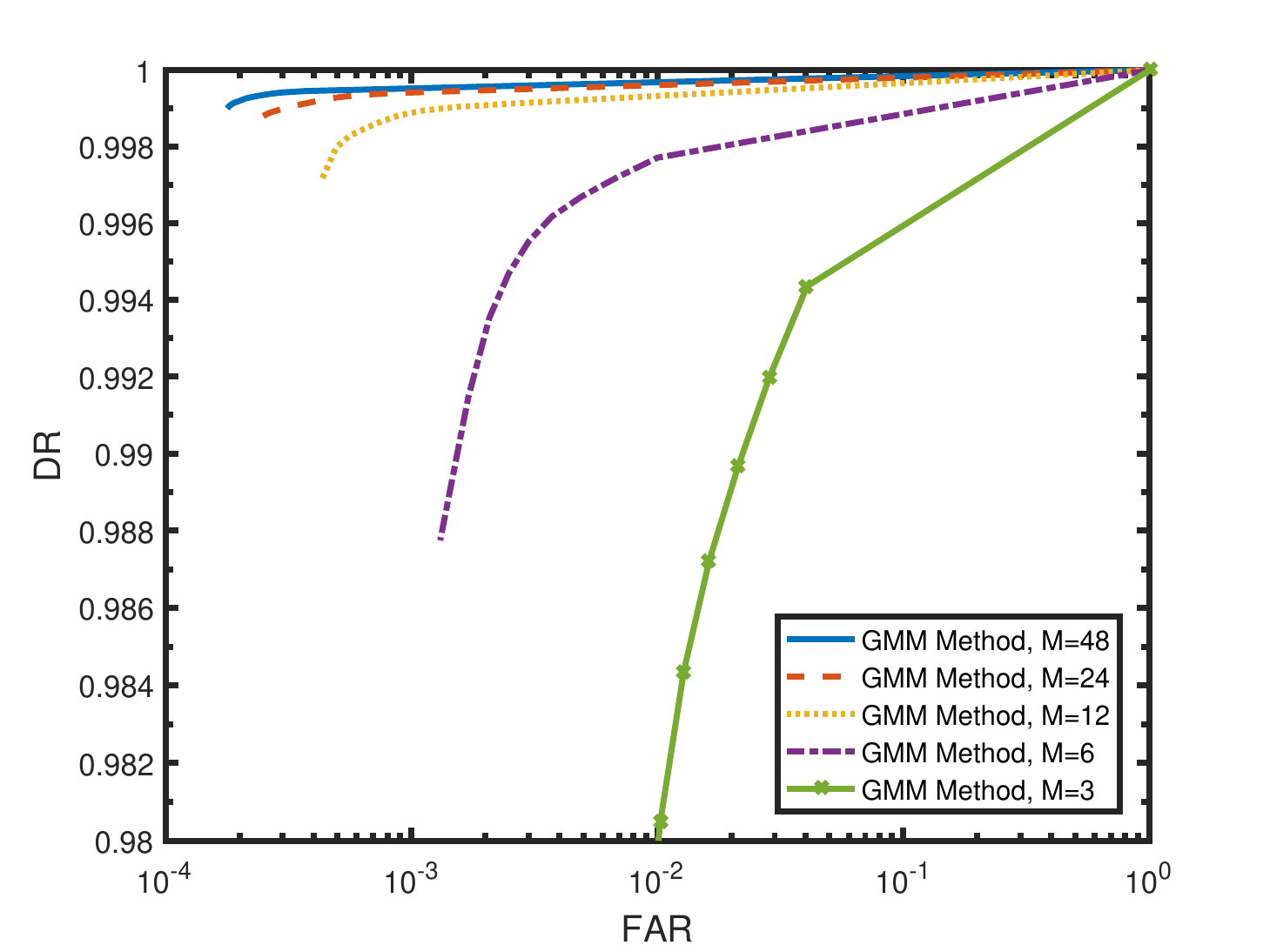}%
\label{roc_M_2}}
\caption{ROC curves for different values of $M$}
\label{roc_M}
\end{figure}

%%% 1000 data per block
%%% 99 data blocks
%%% 1 train block
%%% 50% AI
%%% text from old figure
Fig. \ref{roc_M} shows the influence of the accuracy of the channel estimation in form of different amounts of equally spaced estimated points in frequency domain, ranging from $M=48$ (each active data carrier considered) to $M=3$ (every $16$-th active data carrier considered). In Fig. \ref{roc_M_1}, the respective ROC curves are shown in linear scale and Fig. \ref{roc_M_2} shows them in logarithmic scale. In case of $M=3$, the detection rate is $97.97\%$ at a false alarm rate of $1\%$, while in case of the channel estimation with $48$ estimated bins, the detection rate is at $99.98\%$ at the same false alarm rate. For other values of $M$, the detection rate is at $99.77\%$, $99.93\%$ and $99.96\%$ for $M=6$, $M=12$ and $M=24$ respectively.
%Fig. \ref{training} shows the performance of the GMM based clustering approach dependent on the amount of training data used for initialization of the model at low attack intensities. In Fig. \ref{training 1} the results are shown for a feature space of $M=24$. For $AI=2.5\%$ and $10$ training blocks of size $1000$ the detection rate is $92.14\%$ at a false alarm rate of $0.35\%$. For $AI=5\%$, $P_{\mbox{\scriptsize D}}$ is at $99.88\%$ while $P_{\mbox{\scriptsize FA}}$ is at $0.38\%$. In Fig. \ref{training 2} the feature space is $M=12$. If $AI=2.5\%$, the detection rate results in $98.22\%$ at a false alarm rate of $1.13\%$. In case of $AI=5\%$, the detection rate is at $99.3\%$ at a false alarm rate of $1.57\%$.

\subsection{Discussion of Results}
The experimental evaluation of our method shows, that the performance of our GMM based clustering method increases as the feature space, which is in our case the number of estimated subcarrier per message $M$, also increases. If a false alarm rate of $1\%$ is considered, the performance gain in case of $M=48$ is $2.05\%$ compared to $M=3$. Additionally, we proved that the GMM method performs better than the MSE detection method. Here, in case of a false alarm rate of $5.83\%$, the performance gain of the GMM over MSE method is $23.37\%$.

%%% 1000 data per block
%%% 70 data blocks
%%% 48 sc est.
%%% 5% AI

\section{Conclusion and Future Work}
\label{CONC}
%Due to security flaws in today commonly applied protocols and systems in industrial scenarios, it is essential for future systems to provide adequate security measures which rely on both, conventional cryptography, as well as consideration of unique physical conditions of the present environment. Consequently, 
Our proposed method of taking characteristics of the physical layer into account in order to identify and authenticate transmitters of MC-MTC messages seems to be a promising technique in order to achieve that goal in a very efficient way. The combination of both, MC-MTC and clustering of channel estimates is essential considering system efficiency, as both rely on frequent transmission of data packets. We can reuse channel estimations and make decisions on the authenticity of received data packets with little effort.
Though the maximum achieved detection rate is high at $99.98\%$, the method needs still to be improved in order to get more reliable decisions. To gain robustness due to errors in channel estimations induced by noise, approaches such as in \cite{Ambekar.2012b} might be suited in order to reduce this effect based on pre-processing of channel estimates. Another issue that needs to be investigated is the amount of training data used in order to initialize the model, as well as the assumed attack intensity. Additionally we also want to focus on a mobile setup with little to moderate velocities in order to verify, that the method also works well under these conditions.

\section*{Acknowledgment}
A part of this work has been supported by the Federal Ministry of Education and Research of the Federal Republic of Germany (BMBF) in the framework of the project 16KIS0267 HiFlecs. The authors would like to acknowledge the contributions of their colleagues, although the authors alone are responsible for the content of the paper which does not necessarily represent the project.



\bibliographystyle{IEEEtran}
 %argument is your BibTeX string definitions and bibliography database(s)
\bibliography{references_v1}

%\begin{thebibliography}
%\printbibliography
%
%\bibitem{Que1}
%Reference 1\\
%
%
%TBD (all): approx. 1/4 page

\end{document}